\documentclass[runningheads]{llncs}
\usepackage[T1]{fontenc}
\usepackage{amsmath,epsfig}
\usepackage{amssymb,booktabs,bbding,pifont,subfigure}
\usepackage{epstopdf,graphicx}
\usepackage{cite}
\usepackage[citecolor=green]{hyperref}
\begin{document}
%

\title{Full-Frequency Dynamic Convolution: a Physical Frequency-Dependent Convolution for Sound Event Detection}
\titlerunning{Full-Frequency Dynamic Convolution}
%

\author{Haobo Yue \and
Zhicheng Zhang\thanks{Zhang is the corresponding author.} \and Da Mu \and Yonghao Dang \and Jianqin Yin \and Jin Tang}
\authorrunning{Yue et al.}
%
\institute{Beijing University of Posts and Telecommunications, Beijing, China\\
\email{\{hby, zczhang\}@bupt.edu.cn}}
\maketitle              
%
%
\begin{abstract}
Recently, 2D convolution has been found unqualified in sound event detection (SED). It enforces translation equivariance on sound events along frequency axis, which is not a shift-invariant dimension. To address this issue, dynamic convolution is used to model the frequency dependency of sound events. In this paper, we proposed the first full-dynamic method named \emph{full-frequency dynamic convolution} (FFDConv). FFDConv generates frequency kernels for every frequency band, which is designed directly in the structure for frequency-dependent modeling. It physically furnished 2D convolution with the capability of frequency-dependent modeling. FFDConv outperforms not only the baseline by 6.6\% in DESED real validation dataset in terms of PSDS1, but outperforms the other full-dynamic methods. In addition, by visualizing features of sound events, we observed that FFDConv could effectively extract coherent features in specific frequency bands, consistent with the vocal continuity of sound events. This proves that FFDConv has great frequency-dependent perception ability. 
%
\end{abstract}
\section{Introduction}\label{intro}
Sound event detection (SED) is one of the subtasks of computational auditory scene analysis (CASA)~\cite{Rouat08}, which helps machines understand the content of an audio scene. Similar to visual object detection~\cite{Zhengxia23} and segmentation~\cite{Jiaxing20}, SED aims to detect sound events and corresponding timestamps (onset and offset), considered as a prior task of automatic speech recognition (ASR) and speaker verification. It has wide applications in information retrieval~\cite{jin12}, smart homes~\cite{Debes16}, and smart cities~\cite{Bello18}. 

Initially, methods from other domains have greatly promoted the development of SED. Some methods from computer vision, such as SENet~\cite{Hu18}, SKNet~\cite{Li19}, and CBAM~\cite{Woo18} improved the capacity of feature representation of the network, adding the attention mechanism to the convolution network. The method from speech processing, for example, conformer~\cite{Na21} also improved the representation capacity, which added the local modeling to the transformer structure by inserting a convolution layer. These methods didn't study the characteristics of audio data,  resulting in not great detection performance. Specifically, SENet, SKNet, and CBAM are designed on image data with a clear 2D spatial concept, while audio data is a time sequence. Conformer is designed on speech data containing only the speech sound event, meaning time-frequency patterns of speech data are distributed only in a certain fixed frequency band. However, audio data always contains multiple sound events, and so has diverse time-frequency patterns of sound events.

\begin{figure}[t]
\centering
\includegraphics[scale=0.52]{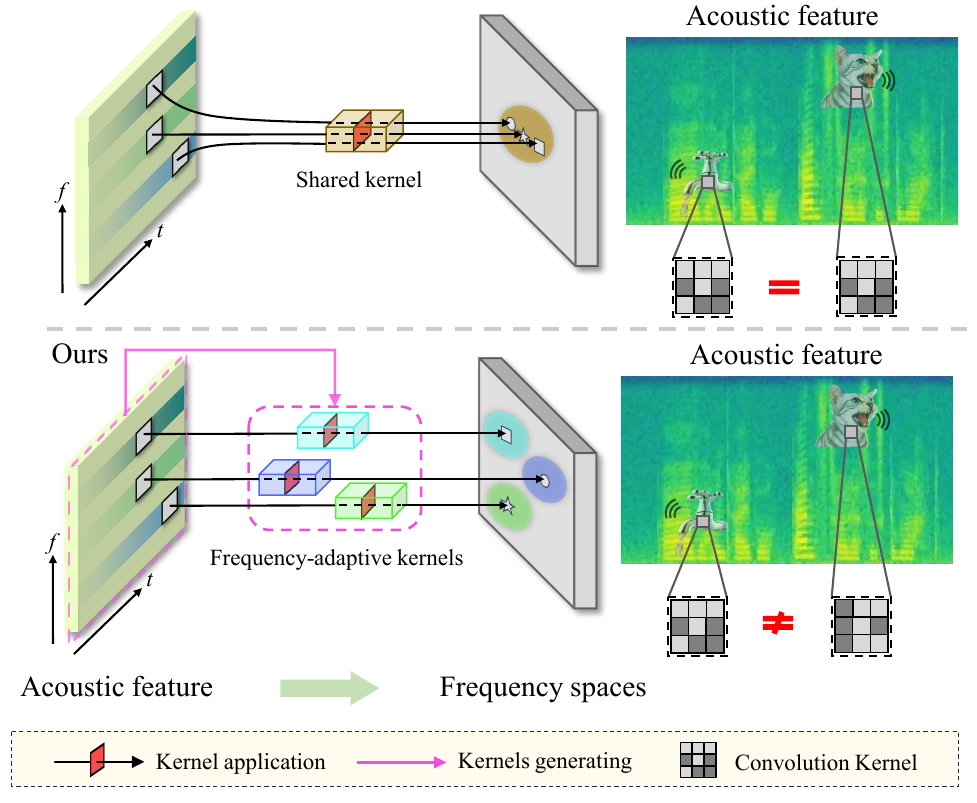}
\caption{Illustration of frequency-dependent modeling. Top models time-frequency patterns in the same space with a shared kernel. Bottom models them in serval spaces with frequency-adaptive kernels, in which time-frequency patterns specific to sound events can be considered.}
\label{fig:introduction}
\end{figure}

Recently, the characteristics of audio data started to be studied, and the dynamic convolution network has been tried in SED. Dynamic convolution network~\cite{xu16} was initially proposed for video prediction. It was designed to generate future frames based on the motion pattern within a particular video. The parameters of the dynamic convolution kernel are always adapted to the input. In SED, different sound events are distributed in different frequency regions, and this frequency dependence is invariant over time. This has motivated some researchers to investigate whether the adaptivity of dynamic convolution can improve the capability of 2D convolution in modeling the frequency dependence of sound events.~\cite{nam22} proposed frequency dynamic convolution (FDConv), which found that the time-frequency spectrogram is not translation invariant on frequency dimension like image data. FDConv extracts frequency-adaptive attention weights from input for several pre-initialized convolution kernels. These kernels are then weightedly combined in the number dimension to obtain one convolution kernel. Then, the combined kernel is convoluted with the input in a standard manner. ~\cite{Shengchang23} proposed multi-dimensional frequency dynamic convolution (MFDConv), which extends the frequency-adaptive dynamic properties of convolutional kernels to more dimensions of the kernel space, i.e. in-channels, out-channels, and kernel numbers.

Although FDConv and MFDConv have achieved great performance, they are essentially the same as basic convolution, which is spatially shared. They belong to semi-dynamic convolution in the field of dynamic convolution. As shown in the upper part of Fig.~\ref{fig:introduction}, their perception abilities of different frequency bands are identical. They can only model time-frequency patterns in one representation space, where sound events are not easily recognized from each other. Compared with semi-dynamic convolution, full-dynamic convolution~\cite{xu16,Esquivel19,Zhi20,Xinlong20,Zhou21} attracts more attention recently, which uses a separate network branch to predict a specific filter for each pixel.~\cite{Zhou21} found this type of dynamic convolution is equivalent to applying attention on unfolded input features, which enables it more effective when modeling complex patterns. Sound events' time-frequency patterns are highly frequency-dependent, and full-dynamic convolution can model features of spatial pixels with different filters. Full-dynamic convolution may be optimal in dealing with recognizing sound events. 

In this paper, we propose a novel method named \emph{full-frequency dynamic convolution} (FFDConv), which is the first full-dynamic convolution method for SED. As shown in the lower part of Fig.~\ref{fig:introduction}, FFDConv generates frequency-specific kernels, resulting in distinct frequency representation spaces. This design is applied directly in the network structure for frequency-dependent modeling. In this way, the 2D convolution is physically furnished with the capability of frequency-dependent modeling, so that the specific time-frequency patterns can be acquired for different sound events. In the end, sound events can be easily recognized from each other in subsequent classification. 

\noindent \textbf{Contributions.} 
(1) We proposed full-frequency dynamic convolution that can model time-frequency patterns in independent representation spaces. This method will extract more discriminative features of sound events, resulting in effective classification.
(2) The Proposed method outperforms not only baseline but also pre-existing full dynamic filters method in other domain.
(3) By visualizing features of sound events, we found the ability to model temporally coherent features is essential to the detection of sound events. And the FFDConv has this ability.

\section{Related Work}
Recently, sound event detection has achieved great success with the help of deep learning. Existing methods include uninitialized learning and fine-tuning pretrained models.

\noindent\textbf{Uninitialized Learning in SED.} Most uninitialized Learning methods employ convolution networks. They either use initial version networks from the computer vision domain or design a new convolution network. As for methods from the computer vision domain, viewing the audio spectrogram as 2D image data, SENet~\cite{Hu18}, SKNet~\cite{Li19}, and CBAM~\cite{Woo18} directly extract features from the audio spectrograms, not considering the physical consistency between standard convolution and audio spectrogram. As for designing a new convolution network, finding the audio spectrogram is not translation invariant on frequency dimension like image data, FDConv~\cite{nam22} designed a frequency-dependent convolution, which equipped the basis convolution with the capacity to model frequency dependence of sound events. To further improve this capacity, MFDConv~\cite{Shengchang23} extended convolutional kernels' frequency-adaptive dynamic properties to more kernel space dimensions, i.e., in-channels, out-channels, and kernel numbers.

\noindent\textbf{Fine-tuning Pretrained Models in SED.} 
In comparison, fine-tuning pretrained models always initialize the networks with weights from the upstream tasks. Chosen models either come from out-of-domain task (vision pre-training) or in-domain task (audio pre-training). As for fine-tuning models from the audio pre-training task, AST-SED~\cite{li23_ICASSP} and PaSST-SED~\cite{li23_interspeech} fine-tuned the audio spectrogram transformer(AST)~\cite{gong21_interspeech} with task-aware adapters in SED. ATST-SED~\cite{shao23_arxiv} fine-tuned the ATST~\cite{li24_TASLP} and BEATs~\cite{chen23_ICML} with a two-stages training strategy. As for fine-tuning models from the vision pre-training task, HTS-AT~\cite{chen22_ICASSP} fine-tuned the swin-transformer~\cite{Liu21_ICCV} in the SED task.

\section{Mthodology}\label{sec: method}
\subsection{Full-Dynamic Convolution}
A basic 2D convolution can be denoted as $y = \boldsymbol{W} \ast x + \boldsymbol{b}$, where $x\in{\mathbb{R}^{T \times F \times C_{in}}}$ and $y\in{\mathbb{R}^{T \times F \times C_{out}}}$ denote the input feature and output feature; $\boldsymbol{b} \in {\mathbb{R}^{C_{out}}}$ and $\boldsymbol{W} \in {\mathbb{R}^{k \times k \times C_{in} \times C_{out}}}$ denote the bias and weight of a basic convolution kernel. In contrast to basis convolution, full-dynamic convolution ~\cite{xu16} leverages separate network branches to generate the filters for each pixel. Full-dynamic convolution operation can be written as:
\begin{equation}\label{eqn:1}
    \begin{aligned}
        y = \boldsymbol{Concat}&{(\boldsymbol{W}_{t,f} \ast x(t,f))} \\
         \boldsymbol{W}_{t,f} =\;&\boldsymbol{G}(x,t,f)
    \end{aligned}
\end{equation}
where $\boldsymbol{W}_{t,f}$ denotes weights of filter for the current pixel; The $\boldsymbol{G}$ is the filter generating function; $\boldsymbol{Concat}$ here aims to convey that convolution operation of each pixel is independent and parallel. For simplicity, the bias term is omitted.

\subsection{Overall of Proposed Method}
As is commonly understood, different sound events have different frequency band distributions. For instance, catcall, which is sharp, shrill, and high-pitched, is often heard in the high-frequency range; running water, which is low, soft, and soothing, is often heard in the low-frequency range. Based on this, we explore designing a new convolution for SED, which can capture the distribution of frequency bands and model time-frequency patterns of sound events in different frequency representation spaces.

\begin{figure}[t]
    \centering
    \includegraphics[scale=0.37]{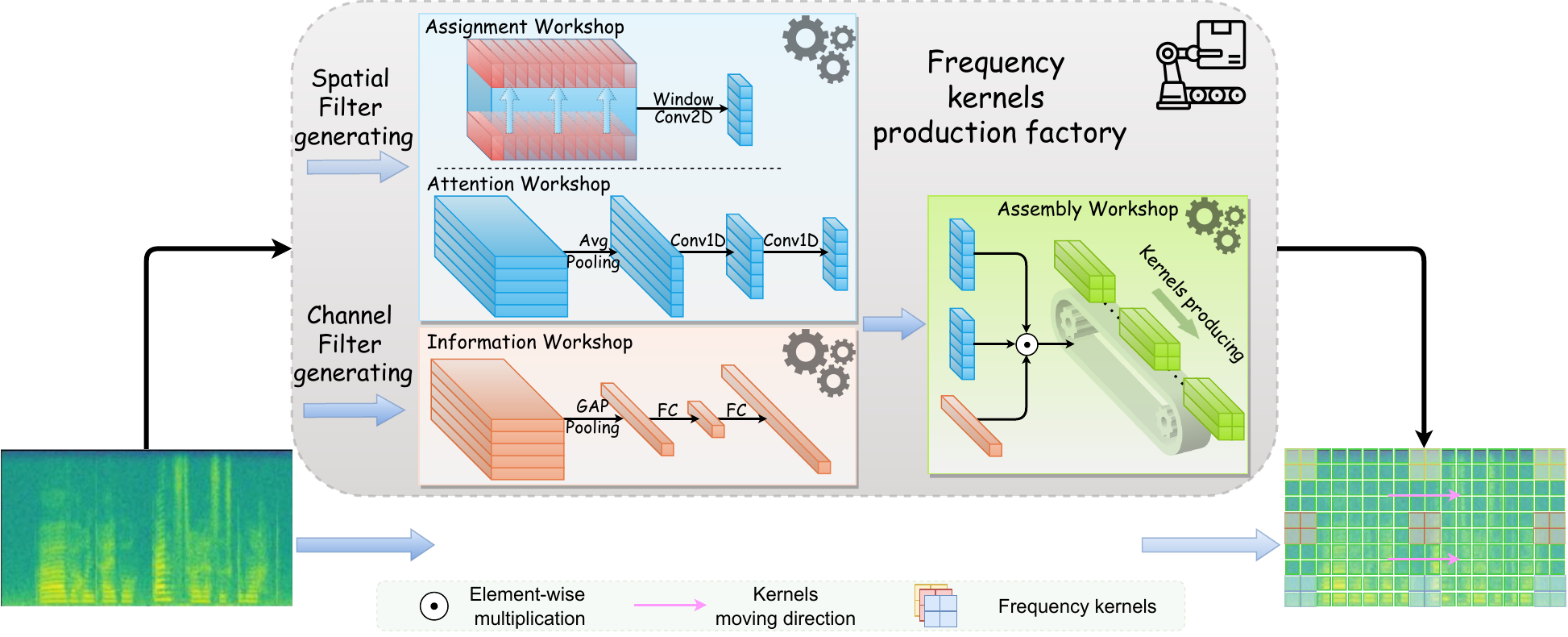}
    \caption{Illustration of full-frequency dynamic convolution. In general, the factory produces frequency-dependent kernels from acoustic feature, and then kernels are convoluted with input along the time axis. In the factory, there are two workshops aiming to produce spatial filters and channel filters, respectively. And they are integrated in the assembly workshop.}
    \label{fig:method_a}
\end{figure}

Inspired by full dynamic convolution~\cite{Zhou21}, we designed the full-frequency dynamic convolution (FFDConv) for SED. Overall, as shown in Fig.~\ref{fig:method_a}, FFDConv employs a separate branch to predict kernels for each frequency band, in which the content of kernels is based on input feature. In the kernel-generating branch, there are two sub-branches: the spatial filter-generating branch for the spatial space of kernels and the channel filter-generating branch for the channel space of kernels. After spatial and channel filters are obtained, they are combined and then convoluted with the input feature. Note that similarly, full-temporal dynamic convolution (FTDConv) predicts kernels for each temporal frame, and kernels are convoluted with input along the frequency axis.

\subsection{Full-Frequency Dynamic Convolution}
Unlike the previous semi-dynamic convolution, FFDConv is designed directly in the structure for frequency-dependent modeling. It models the feature along the frequency axis in different representation spaces. Mathematically, FFDConv can be written as:
\begin{equation}\label{eqn:2}
    \begin{aligned}
        y = \boldsymbol{Concat} & {(\boldsymbol{W}_{f} \ast x(f) , \, dim=f)} \\
         \boldsymbol{W}_{f} =&\; G_{s}(x,f) \odot G_{c}(x,f)
    \end{aligned}
\end{equation}
where $\boldsymbol{W}_f$ is the content-adaptive kernel for the $f^{th}$ frequency band; $x(f) \in \mathbb{R}^{T}$ is the $f^{th}$ frequency band of input feature; $\boldsymbol{G}_{s}$ and $\boldsymbol{G}_{c}$ are the spatial and channel filter-generating function; $\odot$ denotes the elemental dot product operator. For clarity, $\boldsymbol{Concat}$ here aims to convey that $\boldsymbol{W}_f$ is convoluted with input along the time axis and the operation is parallel.

\begin{figure}[t]
    \centering
    \includegraphics[scale=0.62]{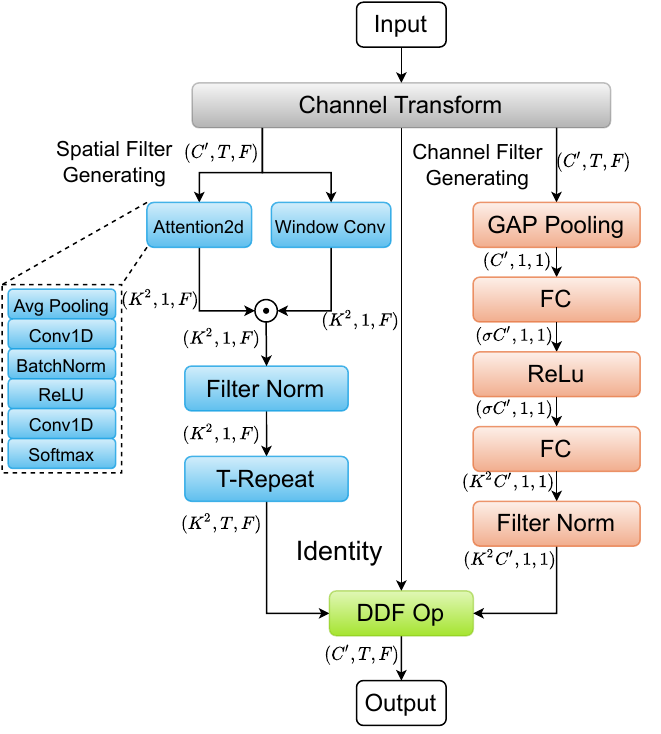}
    \caption{Details of the FFDConv}
    \label{fig:method_b}
\end{figure}
FFDConv employs a separate branch to generate convolution kernels for each frequency band, in which there are two sub-branches: spatial filter-generating branch and channel filter-generating branch. The spatial filter-generating module is designed to predict the spatial content of dynamic kernels, and the channel-generating module is designed to predict the channel content of dynamic kernels. For efficiency, the dynamic filters are decoupled into spatial and channel ones, following~\cite{Zhou21}.

\noindent \textbf{Spatial Filter Generating.} As illustrated in Fig.~\ref{fig:method_b}, we use a standard Conv2D to compress the time dimension of input and map channel dimension from $C$ to $K^2$, whose kernel weight $W \in \mathbf{R}^{C \times K^2 \times T \times W}$, where $W$ is the window size of the kernel in the frequency dimension. It moves along the frequency axis when convoluted with input. In this way, not only are the adjacent frequency components considered, but information along the time axis is aggregated. Then, the spatial filter of FFDConv is obtained, which assigns $K \times K$ spatial weight to every frequency kernel and is highly related to the input. Consequently, FFDConv can model features from different frequency bands of the input in independent representation spaces. In comparison, the convolution map in FTDConv is $W \in \mathbf{R}^{C \times K^2 \times W \times F}$, and it moves along the time axis when convoluted with input, resulting in $K \times K$ spatial weight to every time kernel. The convolution map in full dynamic convolution~\cite{Zhou21} is $W \in \mathbf{R}^{C \times K^2 \times 1 \times 1}$, and assigns $K \times K$ spatial weight to every pixel finally.

Considering these representation spaces may be far apart from each other, we employ an attention2d module following~\cite{nam22} to limit individual differences between them so as not to be too large. Finally, the spatial filter is passed through a Filter-Norm module following~\cite{Zhou21}, avoiding the gradient vanishing/exploding during training.

\noindent \textbf{Channel Filter Generating.} As illustrated in Fig.~\ref{fig:method_b}, the channel filter generating module is similar to the SE block~\cite{Hu18}. It compresses the time and frequency feature of input by applying an average pooling and maps the channel dimension from $C$ to $CK^2$ by two fully connected (FC) layers. Between two fully connected layers, the ReLU activation function is applied to introduce non-linearity. After input is passed through this module, the channel filter of FFDConv is obtained, which assigns $C$ channel weight to each spatial location of the frequency kernel. It should be noted that the channel filter for $F$ frequency kernels is the same. In the end, the channel filter is also passed through the Filter-Norm~\cite{Zhou21}. The spatial and channel filters are mixed by dot product, and the full frequency kernels are obtained. We then use them to model time-frequency patterns of input features. Note that full dynamic convolution~\cite{Zhou21} and FTDConv have the same channel filter generating branch.

\subsection{FFDConv Block}
Considering that the frequency kernels of FFDConv don't have the ability to change the channel dimension of input features, we design an FFDConv block that contains the channel mapping. As illustrated in Fig.~\ref{fig:method_b}, firstly, the channel dimension of input is mapped from $C_{in}$ to $C_{out}$ after passing through the channel transformation module, where a standard 2D convolution is employed. Then, based on the input feature, the spatial and channel filters are obtained by passing through the spatial and channel filter generating module. Full-frequency dynamic kernels are obtained by mixing the spatial and channel filters. Finally, the kernels are convoluted with input along the time axis.

In the actual algorithm, following~\cite{Zhou21}, spatial filters, channel filters, and input are sent to DDF operation to get the output, which is implemented in CUDA, alleviating any need to save intermediate multiplied filters during network training and inference. Note that the DDF op needs $H \times W$ spatial filters. We repeat the $1 \times F$ spatial filters to $T \times F$ so that the kernel's weights are the same along the time axis when convoluted with input in $f^{th}$ frequency band.

\section{Experiment}\label{sec:exp}
\subsection{Dataset, Metrics and Implementation Details}
\noindent \textbf{Dataset.}All experiments are conducted on the dataset of Task 4 in the DCASE 2022. The training set consists of three types of data: weakly labeled data (1578 clips), synthetic strongly labeled data (10000 clips), and unlabeled in-domain data (14412 clips). The real validation set (1168 clips) is used for evaluation. The input acoustic feature is the log Mel spectrogram extracted from 10-second-long audio data with a sampling rate of 16 kHz. The feature configuration is the same as [13], in which the input feature has 626 frames and 128 mel frequency bands.

\noindent \textbf{Implementation Details.} The baseline model is the CRNN architecture~\cite{Emre17}, which consists of 7 layers of conv blocks and 2 layers of Bi-GRU. Attention pooling module is added at the last FC layer for joint training of weakly labeled data, and mean teacher (MT)~\cite{Tarvainen17} is applied for consistency training with unlabeled data for semi-supervised learning. Data augmentations such as MixUp~\cite{zhang18}, time masking~\cite{park19}, frame-shift, and FilterAugment~\cite{Hyeonuk22} are used. The data augmentation parameters are identical to~\cite{nam22}.The metrics hyper-parameters are identical to~\cite{nam22}. The model is trained using the Adam optimizer with a maximum learning rate of 0.001, and ramp-up is used for the first 80 epochs.

\noindent \textbf{Metrics.} Poly-phonic sound event detection scores (PSDS)~\cite{Bilen20}, collar-based F1 score (EB-F1)~\cite{Mesaros16}, intersection-based F1 score (IB-F1)~\cite{Bilen20} are used to evaluate the model performance. Median filters with fixed time length are used for post-processing, and sound events have different thresholds from each other to obtain hard predictions for calculating EB-F1. These metrics have different focuses. PSDS1 and CB-F1 reflect more on the system's capacity for detecting sound events. PSDS2 and IB-F1 reflect more on the system's capacity for classifying and differing sound events.

\begin{table}[t]
    \centering 
    \setlength{\abovecaptionskip}{3pt}
    \setlength{\belowcaptionskip}{10pt}
    \setlength{\tabcolsep}{4pt}
    \caption{SED performance comparison between models using different convolutions on the real validation set. The best results are in \textbf{bold}, and the second best are in \underline{underlined}. * denotes the results from our implementation using the codebase from~\cite{nam22}.}
    \begin{tabular}{c c c c c c}
        \bottomrule
        \textbf{Model} & \textbf{Params} & \textbf{PSDS1} $\uparrow$ & \textbf{PSDS2} $\uparrow$ & \textbf{CB-F1} $\uparrow$ & \textbf{IB-F1} $\uparrow$ \\
        \hline
        Baseline*~\cite{Emre17} & 4M & 0.370 & 0.579 & 0.469 & 0.714 \\
        \hline
        DDFConv*~\cite{Zhou21} & 7M & 0.387 & 0.624 & 0.467 & 0.720 \\
        FTDConv* & 7M & 0.395 & 0.651 & 0.495 & \underline{0.740} \\
        SKConv~\cite{Zheng21_ICASSP} & -- &  0.400 & -- & 0.520 & -- \\
        FDConv*~\cite{nam22} & 11M & 0.431 & 0.663 & 0.521 & 0.738 \\
        MFDConv~\cite{Shengchang23} & 33M & \textbf{0.461} & \underline{0.680} & \textbf{0.542} & -- \\
        FFDConv* & 11M & \underline{0.436} & \textbf{0.685} & \underline{0.526} & \textbf{0.751} \\
        \toprule
    \end{tabular}
    \label{tab:table1}
\end{table}
\subsection{Full-Frequency Dynamic Convolution on SED}
We compared the performances of baseline with full dynamic convolution methods, including decoupled dynamic convolution (DDFConv)~\cite{Zhou21}, full-temporal dynamic convolution (FTDConv), and full-frequency dynamic convolution (FFDConv). For full dynamic convolution methods, dynamic convolution layers replaced all convolution layers except the first layer from the baseline model~\cite{Emre17}. Besides, some typical convolution methods are also compared.

Compared with dynamic convolution methods, from Table~\ref{tab:table1}, three types of full dynamic convolution can all outperform the baseline, which proves full dynamic convolution qualifies in SED. In addition, it can be seen that the effects of three types of convolution are in increasing order. First, FTDConv and FFDConv employ content-adaptive temporal or frequency kernels, which can be viewed as giving prior knowledge to SED compared with DDFConv. Second, FFDConv outperforms FTDConv, which can prove that time-frequency patterns of sound events are highly frequency-dependent, and this dependency is time-invariant. Moreover, FFDConv models acoustic features with different kernels along the frequency axis, which can be thought to be frequency components modeled in different representation spaces. As if components of the feature are split into different frequency spaces and then reassembled. This is consistent with the characteristics of sound events.

Compared with other typical methods, such as MFDConv~\cite{Shengchang23}, SKConv~\cite{Zheng21_ICASSP}, FDConv~\cite{nam22}, FFDConv still demonstrates competitive performances. Especially in terms of PSDS2 and IB-F1, FFDConv was the best among all convolution methods. It approves the effectiveness of FFDConv, which captures the information of different frequency band distributions for different sound events and then models more differentiated time-frequency patterns for them, favoring the classification of sound events. In terms of the PSDS1 and CB-F1, the FFDConv is suboptimal and slightly higher than FDConv~\cite{nam22}. FFDConv's kernels are time-invariant in some frequency band, the same as FDConvs~\cite{nam22}, leading to a close result. Compared to the MFDConv~\cite{Shengchang23}, the latest state-of-the-art convolution method, FFDConv can also get competitive performances, with a 66.7\% decrease in the number of parameters. 

\begin{figure}[t]
    \centering
    \includegraphics[scale=0.22]{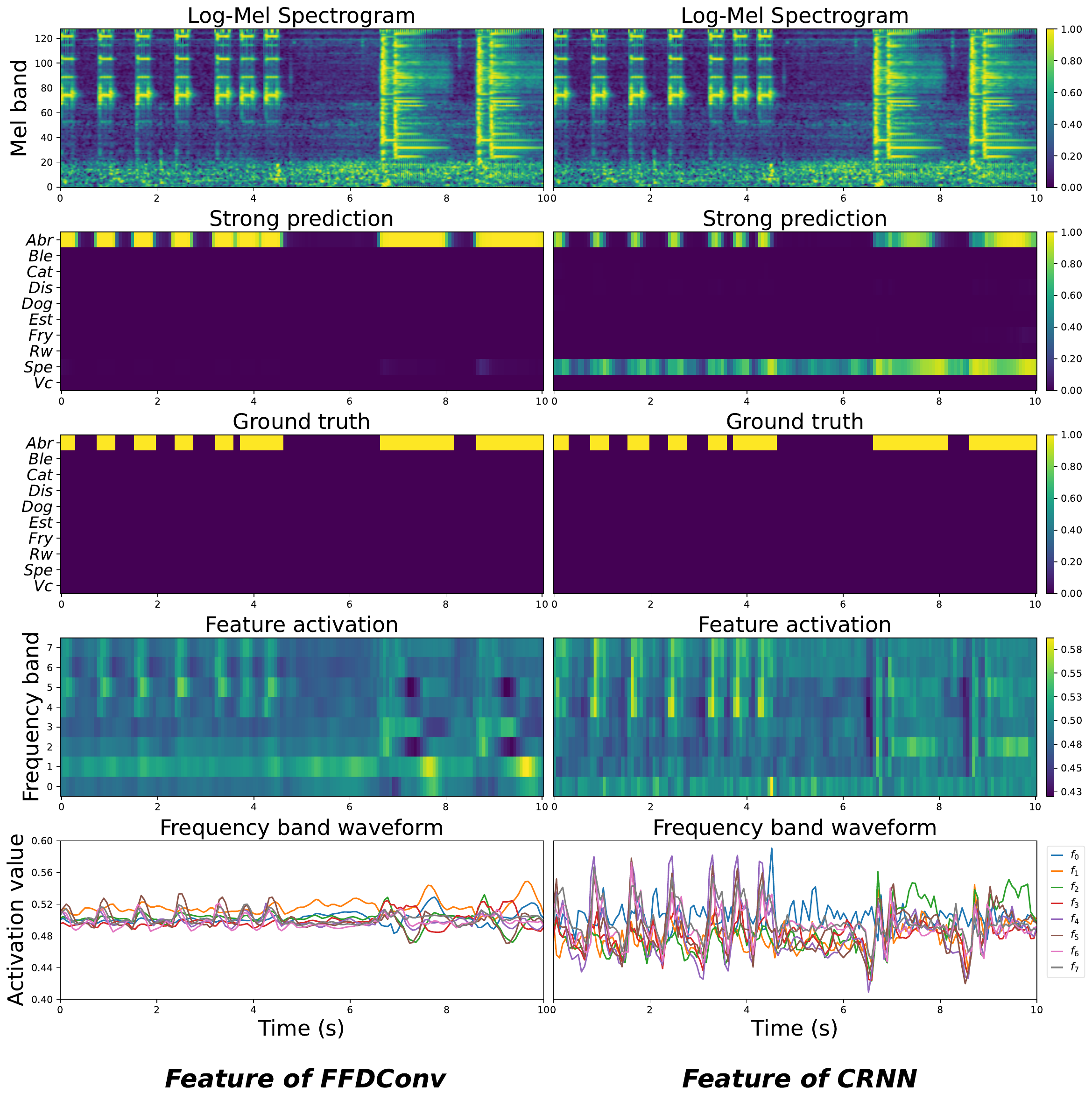}
    \caption{Feature comparison of FFDConv and CRNN. Features activation of the 5th Conv block are shown in the 4th row. The trends of frequency band features over time are shown in the 5th row. Note that y-axis labels of strong prediction are abbreviations of the sound event categories. For example, Abr stands for Alarm bell ringing.}
    \label{fig:feature}
\end{figure}
\subsection{Fine-Grained Modeling Study}
To explore FFDConv's ability to understand acoustic spectral information at a fine-grained level. We visualized feature of the middle layer. More visualizations can be found in the supplementary material.

The visualization results are shown in Fig.~\ref{fig:feature}. Comparing the features of FFDConv and CRNN, we can see that most of the time-frequency patterns modeled by CRNN are temporally isolated and disjoint. In contrast, FFDConv's patterns and their neighbors are in a whole, thereby forming a distinct time-frequency representation. Moreover, this phenomenon can also be found in trends of frequency band features over time. The waveforms of FFDConv are smoother than CRNN. Specifically, the duration of peak and trough is longer in FFDConv's waveform, which results from the feature being mostly coherent over time. There are more pulses in the resting state of CRNN's waveforms, which are in a disorganized state. Besides, the distributions of frequency band features are consistent with alarm\_bell\_ring's spectrogram in FFDConv's waveforms. The values of the low-frequency band features are smaller than those of the middle and high-frequency bands when the alarm bell rings. However, the differences between frequency bands in CRNN are ambiguous. As for the model's prediction, the CRNN's isolated features directly lead to the incoherent output compared with ground truth, which proves that the feature's coherence over time is essential. Interestingly, the low-frequency white noise of the sound clip is filtered by FFDConv, but CRNN tagged it as speech. This has to do with that dynamic convolution concentrates more on high-frequency texture information, and white noise in the spectrogram lacks clear contour information.

Most SED models are trained in a frame-based supervised way, which always leads to the feature and output being discrete over time. However, FFDConv can alleviate this by frequency-dependent modeling, which models different patterns for frequency bands, leading to a distinct representation of a sound event. This modeling way is like an attention mechanism in which the distribution of frequency band information of the spectrogram is maintained. Besides, the convolution kernel for a frequency band is shared in all frames, which produces temporally coherent representations. This is consistent with both the continuity of the sound waveform and the vocal continuity of sound events.

\subsection{Ablation Study}
We compared the performance of different window sizes of the build kernel when generating spatial filters. Note that the size of the spatial filter $K$ is set to 3. 
\begin{table}[t]
    \centering 
    \setlength{\abovecaptionskip}{3pt}
    \setlength{\belowcaptionskip}{10pt}
    \caption{\emph{Comparison of different window size, W.}}
    \begin{tabular}{c c c c c}
        \bottomrule
        \textbf{Model} & $Atten$ & $W$ & \textbf{PSDS1} & \textbf{PSDS2} \\
        \hline
                & \ding{55} & 3 & 0.421 & 0.650 \\
                & \checkmark & 1 & 0.421 & 0.659 \\
        FFDConv & \checkmark & 3 & \textbf{0.436} & \textbf{0.685} \\
                & \checkmark & 5 & 0.423 & 0.656 \\
                & \checkmark & 7 & 0.432 & 0.666 \\
        \toprule
    \end{tabular}
    \label{tab:table2}
\end{table}

The results are shown in Table~\ref{tab:table2}. With constraints of the attention module, FFDConv can get better performance. This proves that before attention, spatial filters of different frequency spaces may have a large distance from each other. The performance of FFDConv is the best when window size is set to 3. This is because the adjacent frequency components are considered compared to size 1 when generating the spatial filter, and size 5 may suffer from overfitting. In addition, it's interesting that the performance recovers when the window size is set to 7. This may have to do with the fact that dynamic convolutions are relatively unstable.

\section{Conclusions}\label{sec:con}
In this paper, we proposed full-frequency dynamic convolution, the first full-dynamic method for SED. Full-frequency dynamic convolution is designed to model time-frequency patterns in different frequency spaces. This design in structure physically furnished 2D convolution with the capability of frequency-dependent modeling. Experiments on the DESED show that full-frequency dynamic convolution is superior to not only baseline but also other full-dynamic convolutions, which proves FFDConv qualifies in SED. In addition, by visualizing features of sound events, we found that FFDConv can extract temporally coherent features in specific frequency bands, which is consistent with the vocal continuity of sound events. This proves that FFDConv has great frequency-dependent perception ability. In the future, we aim to explore new methods to model vocal continuity of sound events.

\subsubsection{Acknowledgements} This work was supported partly by the National Natural Science Foundation of China (Grant No. 62173045, 62273054), partly by the Fundamental Research Funds for the Central Universities (Grant No. 2020XD-A04-3), and the Natural Science Foundation of Hainan Province (Grant No. 622RC675).
%

%
%
%
\bibliographystyle{splncs04}
\bibliography{ICPR2024}

\begin{thebibliography}{10}
\providecommand{\url}[1]{\texttt{#1}}
\providecommand{\urlprefix}{URL }
\providecommand{\doi}[1]{https://doi.org/#1}

\bibitem{Bello18}
Bello, J., Mydlarz, C., Salamon, J.: Sound analysis in smart cities, pp. 373--397. Springer International Publishing (Sep 2017)

\bibitem{Bilen20}
Bilen, {\c{C}}., Ferroni, G., Tuveri, F., Azcarreta, J., Krstulović, S.: {A Framework for the Robust Evaluation of Sound Event Detection}. In: ICASSP. pp. 61--65 (May 2020)

\bibitem{chen22_ICASSP}
Chen, K., Du, X., Zhu, B., Ma, Z., Berg-Kirkpatrick, T., Dubnov, S.: Hts-at: A hierarchical token-semantic audio transformer for sound classification and detection. In: ICASSP. pp. 646--650 (2022)

\bibitem{chen23_ICML}
Chen, S., Wu, Y., Wang, C., Liu, S., Tompkins, D., Chen, Z., Che, W., Yu, X., Wei, F.: {BEAT}s: Audio pre-training with acoustic tokenizers. In: ICML. pp. 5178--5193 (July 2023)

\bibitem{Debes16}
Debes, C., Merentitis, A., Sukhanov, S., Niessen, M., Frangiadakis, N., Bauer, A.: {Monitoring Activities of Daily Living in Smart Homes: Understanding human behavior}. IEEE SPM  \textbf{33}(2),  81--94 (March 2016)

\bibitem{gong21_interspeech}
Gong, Y., Chung, Y.A., Glass, J.: {AST: Audio Spectrogram Transformer}. In: INTERSPEECH. pp. 571--575 (Aug 2021)

\bibitem{Hu18}
Hu, J., Shen, L., Sun, G.: {Squeeze-and-Excitation Networks}. In: CVPR. pp. 7132--7141 (Jun 2018)

\bibitem{xu16}
Jia, X., De~Brabandere, B., Tuytelaars, T., Gool, L.V.: Dynamic filter networks. In: NEURIPS (Dec 2016)

\bibitem{jin12}
Jin, Q., Schulam, P., Rawat, S., Burger, S., Ding, D., Metze, F.: {Event-based video retrieval using audio}. In: INTERSPEECH. pp. 2085--2088 (Sept 2012)

\bibitem{li23_ICASSP}
Li, K., Song, Y., Dai, L.R., McLoughlin, I., Fang, X., Liu, L.: Ast-sed: An effective sound event detection method based on audio spectrogram transformer. In: ICASSP. pp.~1--5 (Jun 2023)

\bibitem{li23_interspeech}
Li, K., Song, Y., McLoughlin, I., Liu, L., Li, J., Dai, L.R.: {Fine-tuning Audio Spectrogram Transformer with Task-aware Adapters for Sound Event Detection}. In: INTERSPEECH. pp. 291--295 (Aug 2023)

\bibitem{li24_TASLP}
Li, X., Shao, N., Li, X.: Self-supervised audio teacher-student transformer for both clip-level and frame-level tasks. IEEE/ACM TASLP  \textbf{32},  1336--1351 (2024)

\bibitem{Li19}
Li, X., Wang, W., Hu, X., Yang, J.: {Selective Kernel Networks}. In: CVPR. pp. 510--519 (Jun 2019)

\bibitem{Liu21_ICCV}
Liu, Z., Lin, Y., Cao, Y., Hu, H., Wei, Y., Zhang, Z., Lin, S., Guo, B.: Swin transformer: Hierarchical vision transformer using shifted windows. In: ICCV. pp. 10012--10022 (Oct 2021)

\bibitem{Mesaros16}
Mesaros, A., Heittola, T., Virtanen, T.: {Metrics for Polyphonic Sound Event Detection}. AS  \textbf{6}(6) (May 2016)

\bibitem{Na21}
Na, T., Zhang, Q.: Convolutional network with conformer for semi-supervised sound event detection. Tech. rep., DCASE2021 Challenge (July 2021)

\bibitem{nam22}
Nam, H., Kim, S.H., Ko, B.Y., Park, Y.H.: {Frequency Dynamic Convolution: Frequency-Adaptive Pattern Recognition for Sound Event Detection}. In: INTERSPEECH. pp. 2763--2767 (Sept 2022)

\bibitem{Hyeonuk22}
Nam, H., Kim, S.H., Park, Y.H.: {Filteraugment: An Acoustic Environmental Data Augmentation Method}. In: ICASSP. pp. 4308--4312 (May 2022)

\bibitem{park19}
Park, D.S., Chan, W., Zhang, Y., Chiu, C.C., Zoph, B., Cubuk, E.D., Le, Q.V.: {SpecAugment: A Simple Data Augmentation Method for Automatic Speech Recognition}. In: INTERSPEECH. pp. 2613--2617 (Sept 2019)

\bibitem{Rouat08}
Rouat, J.: {Computational Auditory Scene Analysis: Principles, Algorithms, and Applications}. IEEE TNN  \textbf{19}(1),  199--199 (Jan 2008)

\bibitem{shao23_arxiv}
Shao, N., Li, X., Li, X.: Fine-tune the pretrained atst model for sound event detection. ARXIV PREPRINT  (Sept 2023)

\bibitem{Jiaxing20}
Sun, J., Li, Y., Lu, H., Kamiya, T., Serikawa, S.: {Deep Learning for Visual Segmentation: A Review}. In: COMPSAC. pp. 1256--1260 (July 2020)

\bibitem{Tarvainen17}
Tarvainen, A., Valpola, H.: {Mean teachers are better role models: Weight-averaged consistency targets improve semi-supervised deep learning results}. In: NEURIPS (Dec 2017)

\bibitem{Zhi20}
Tian, Z., Shen, C., Chen, H.: {Conditional Convolutions for Instance Segmentation}. In: ECCV. pp. 282--298 (Aug 2020)

\bibitem{Xinlong20}
Wang, X., Zhang, R., Kong, T., Li, L., Shen, C.: {SOLOv2: Dynamic and Fast Instance Segmentation}. In: NEURIPS. pp. 17721--17732 (Dec 2020)

\bibitem{Woo18}
Woo, S., Park, J., Lee, J.Y., Kweon, I.S.: {CBAM: Convolutional Block Attention Module}. In: ECCV. pp. 3--19 (Sept 2018)

\bibitem{Shengchang23}
Xiao, S., Zhang, X., Zhang, P.: {Multi-Dimensional Frequency Dynamic Convolution with Confident Mean Teacher for Sound Event Detection}. In: ICASSP. pp.~1--5 (Jun 2023)

\bibitem{Esquivel19}
Zamora~Esquivel, J., Cruz~Vargas, A., Lopez~Meyer, P., Tickoo, O.: {Adaptive Convolutional Kernels}. In: ICCV Workshops. pp.~0--0 (Oct 2019)

\bibitem{zhang18}
Zhang, H., Cisse, M., Dauphin, Y.N., Lopez-Paz, D.: {mixup: Beyond Empirical Risk Minimization}. ICLR  (Oct 2018)

\bibitem{Zheng21_ICASSP}
Zheng, X., Song, Y., McLoughlin, I., Liu, L., Dai, L.R.: An improved mean teacher based method for large scale weakly labeled semi-supervised sound event detection. In: ICASSP. pp. 356--360 (2021)

\bibitem{Zhou21}
Zhou, J., Jampani, V., Pi, Z., Liu, Q., Yang, M.H.: {Decoupled Dynamic Filter Networks}. In: CVPR. pp. 6647--6656 (Jun 2021)

\bibitem{Zhengxia23}
Zou, Z., Chen, K., Shi, Z., Guo, Y., Ye, J.: {Object Detection in 20 Years: A Survey}. Proc. IEEE  \textbf{111}(3),  257--276 (March 2023)

\bibitem{Emre17}
Çakır, E., Parascandolo, G., Heittola, T., Huttunen, H., Virtanen, T.: {Convolutional Recurrent Neural Networks for Polyphonic Sound Event Detection}. IEEE/ACM TASLP  \textbf{25}(6),  1291--1303 (Jun 2017)

\end{thebibliography}
%





\end{document}